\documentclass[reprint,
superscriptaddress,
 amsmath,amssymb,
 aps,
]{revtex4-1}

\usepackage{graphicx}% Include figure files
\usepackage{dcolumn}% Align table columns on decimal point
\usepackage{bm}% bold math
\usepackage{enumerate}
\usepackage{lipsum}
\usepackage{epstopdf}
\begin{document}

\preprint{APS/123-QED}

\title{Angular Momentum Transfer in Interaction of Laguerre-Gaussian Beams with Atoms and Molecules}% Force line breaks with \\

\author{Pradip Kumar Mondal}
\email{pradip3@phy.iitkgp.ernet.in}
\affiliation{Department of Physics, Indian Institute of Technology Kharagpur, Kharagpur-721302, India.}
\author{Bimalendu Deb}%
 \email{msbd@iacs.res.in}
\affiliation{Department of Materials Science, Indian Association for the Cultivation of Science, Jadavpur, Kolkata 700032, India.}
\author{Sonjoy Majumder}
\affiliation{Department of Physics, Indian Institute of Technology Kharagpur, Kharagpur-721302, India.}

%\affiliation{
%\affiliation{
 %Third institution, the second for Charlie Author
%\affiliation{%



\begin{abstract}
Exchange of orbital angular momentum between Laguerre-Gaussian beam of light and center-of-mass motion of an atom or molecule is well known. We show that orbital angular momentum of light can also be transferred to the internal electronic or rotational motion of an atom or a molecule provided the internal and center-of-mass motions are coupled. However, this transfer does not happen directly to the internal motion, but via center-of-mass motion. If atoms or molecules are cooled down to recoil limit then an exchange of angular momentum between the quantized center-of-mass motion and the internal motion is possible during interaction of cold atoms or molecules with Laguerre-Gaussian beam. The orientation of the exchanged angular momentum is determined by the sign of the winding number of Laguerre-Gaussian beam. We have presented selective results of numerical calculations for the quadrupole transition rates in interaction of Laguerre-Gaussian beam with an atomic Bose-Einstein condensate to illustrate the underlying mechanism of light orbital angular momentum transfer. We discuss how the alignment of diatomic molecules will facilitate to explore the effects of light orbital angular momentum on electronic motion of molecules.
%An article usually includes an abstract, a concise summary of the work
%covered at length in the main body of the article. 
%\begin{description}
%\end{description}
\end{abstract}

\maketitle

\section{Introduction}
Two decades after the pioneering work of Allen and co-workers \cite{allen_92} showing that Laguerre-Gaussian (LG) beams carry well defined orbital angular momentum (OAM), the role of this OAM in interactions of such beams with an atom or a molecule remains an open question. OAM of light usually interacts with the external center-of-mass (c.m.) motion of an atom \cite{enk_94, enk2_94}. In contrast, light polarization which is spin angular momentum of light can interact with the internal electronic motion of an atom. Many researchers have predicted that the field OAM can be transferred to the internal motion of an atom \cite{romero_02, jauregui_04, alex_05, picon2_10, picon_10} or a molecule \cite{babiker_02, alex_06, rury_13} in electronic dipole or quadrupole transitions, while some works \cite{andrews_04, andrews_12} have shown that the field OAM does not interact with molecular chirality. Applications have been proposed based on direct coupling of field OAM to the internal motion \cite{romero_02, veenedaal_07, picon2_10, picon_10, koksal_12, schmi_12, rury_13, afanasev_13} or to the c.m. motion only \cite{muthu_02, lembessis_10, lembessis_13}. The transfer of field OAM to c.m. motion of optically trapped particles (optical spanner effect) \cite{simpson_97, oneil_02, brad_05} and Bose-Einstein condensates (BEC) \cite{andersen_06, wright_08} is well known. But the experiments so far seem to contradict direct coupling of field OAM with internal motion \cite{araoka_05, loffler_11}. It is therefore important to understand how light OAM takes part in light-atom or light-molecule interaction. The OAM of light is associated with the spatial inhomogeneity of field over the beam cross-section. The question we address here: Can an electron in an atom or a molecule feel the spatial variation of the field during its orbital motion?

The paper is organized as follows. In Sec. II, we have developed the theory corresponding to the method of OAM exchange in interaction of LG beam with an atom or a molecule, separately. Sec. III  presents numerical calculations of transition rates in interaction of atomic BEC with LG beam as an example of our theory. Finally, in Sec. IV, we have made some concluding remarks.

\section{Theory}
We consider a LG beam without any off-axis node propagating along $ z $ axis of laboratory frame interacting with a cold atom or a cold molecule whose c.m. wavefunction has extension as large as comparable to the wavelength of the light but smaller than the waist of the beam. The atom or molecule experiences a local field of the type \cite{ alex_05, alex_06, rury_13}

\begin{align}
\textbf{E}(\textbf{r}',t)=\dfrac{\textbf{E}_0}{\sqrt{\vert l\vert !}}\left(\dfrac{{\text{r}'}_{\perp}}{w_0}\right)^{\vert l \vert } \exp (il\phi') \exp [i(kz'-\omega t)],
\end{align} 
where $ {\text{r}'}_{\perp} $ is the projection of $ \textbf{r}' $ on $ xy $ plane. $ l $ and $ w_0 $ are the winding number and the waist of the beam, respectively; and $ \phi' $ is the azimuth. 

\paragraph{\textbf{The atom-radiation interaction:}}
We consider the simplest atomic system composed of a nucleus of positive charge (mass) $ +e $ ( $ m_n $) and an electron of negative charge $ -e $  ( $ m_e $). For simplicity, the spin of the particles is ignored. The center of mass coordinate of the atomic system is $ \textbf{R}_{\text{c.m.}}=(m_e\textbf{r}_e+m_n\textbf{r}_n)/m_t $ with $ m_t=m_e+m_n $ being the total mass, $ \textbf{r}_e $ and $ \textbf{r}_n $ the coordinates of the electron and nucleus, respectively. The Hamiltonian of the atom-field system is  $ H=H_0+H_{\text{I,atom}} $ where $ H_0 $ is unperturbed atomic Hamiltonian and
\begin{align}
H_{\text{I,atom}}= -\int \,d\textbf{r}' \mathcal{P}(\textbf{r}')\cdot \textbf{E}(\textbf{r}',t)+\text{h.c.}
\end{align}
is the interaction Hamiltonian derived in Power-Zineau-Wooley (PZW) scheme \cite{babiker_02, alex_06, cohen_89}. $ \mathcal{P}(\textbf{r}') $ is the electric polarization given by

\begin{align}
\mathcal{P}(\textbf{r}')= -e\frac{m_n}{m_t} \textbf{r} \int_0^1 \,d\lambda \delta (\textbf{r}' - \textbf{R}_{\text{c.m.}}-\lambda \frac{m_n}{m_t}\textbf{r}),
\end{align}
where relative coordinate (internal) $ \textbf{r}=\textbf{r}_e-\textbf{r}_n $. 

The diameter of the region of LG beam of Eq. (1) typically ranges between $ 10^{-4}$ and $10^{-5} $ m \cite{arlt_98, chen_01, sueda_04, andersen_06, li_08, kang_13} while the dimension of an electron orbital in an atom is of the order of a few angstrom. This means $ \vert \textbf{r}\vert \ll \vert  \textbf{R}_{\text{c.m.}}\vert$ and  we therefore use Taylor's expansion 
\begin{widetext}
\begin{align}
\textbf{E}\left( \textbf{R}_{\text{c.m.}}+\lambda \frac{m_n}{m_t}\textbf{r}\right)= \textbf{E}( \textbf{R}_{\text{c.m.}})+\dfrac{\textbf{E}_0}{\sqrt{\vert l\vert !}}\left(\lambda \frac{m_n}{m_t}\right)\left(\dfrac{1}{w_0}\right)^{\vert l\vert} r \left\lbrace{\bf\hat{\textbf{r}}}\cdot\vec{\nabla} \left({\text{r}'}_{\perp}^{\vert l \vert } e^{il\phi'}e^{ikz'}\right)\right\rbrace_{\textbf{R}_{\text{c.m.}}}+\cdots ,
\end{align}
where
\begin{align}
&r \left\lbrace{\bf\hat{\textbf{r}}}\cdot\vec{\nabla} \left({\text{r}'}_{\perp}^{\vert l \vert } e^{il\phi'}e^{ikz'}\right)\right\rbrace_{\textbf{R}_{\text{c.m.}}} \nonumber\\
& = \text{R}_{\text{c.m.}\perp}^{\vert l \vert -1 } e^{il\Phi_{\text{c.m.}}}e^{ ikZ_{\text{c.m.}}}\frac{1}{2}[e^{i\phi}e^{-i\Phi_{\text{c.m.}}}(\vert l\vert +l)+e^{-i\phi}e^{i\Phi_{\text{c.m.}}}(\vert l\vert -l)]r\sin\theta +(ik)\text{R}_{\text{c.m.}\perp}^{\vert l \vert } e^{il\Phi_{\text{c.m.}}}e^{ikZ_{\text{c.m.}}}r\cos\theta.
\end{align}
Substituting Eqs. (3), (4) and (5) into Eq. (2), $ H_{\text{I,atom}} $ can be separated into dipole and quadrupole parts as given by 
\begin{align}
H_{\text{I,atom}}^{d}=\sqrt{\dfrac{4\pi }{3\vert l\vert !}} e\left(\frac{m_n}{m_t}\right) r\sum_{\sigma =0,\pm1}\epsilon_{\sigma}Y_1^{\sigma}({\bf\hat{\textbf{r}}})\left(\dfrac{\text{R}_{\text{c.m.}\perp}}{w_0}\right)^{\vert l \vert } e^{il\Phi_{\text{c.m.}}}e^{ikZ_{\text{c.m.}}} + \text{h.c.}.
\end{align}

\begin{align}
H_{\text{I,atom}}^{q}&=\frac{1}{2}\sqrt{\dfrac{4\pi }{3\vert l\vert !}} e\left(\frac{m_n}{m_t}\right)^2\left(\dfrac{1}{w_0}\right)^{\vert l\vert}r^2\sum_{\sigma =0,\pm1}\epsilon_{\sigma}Y_1^{\sigma}({\bf\hat{\textbf{r}}})\left\lbrace \text{R}_{\text{c.m.}\perp}^{\vert l \vert -1 } e^{il\Phi_{\text{c.m.}}}e^{ ikZ_{\text{c.m.}}}\frac{1}{2}\left[e^{i\phi}e^{-i\Phi_{\text{c.m.}}}(\vert l\vert +l)\right.\right.\nonumber\\
&\left.\left.+e^{-i\phi}e^{i\Phi_{\text{c.m.}}}(\vert l\vert -l)\right]\sin\theta +(ik)\text{R}_{\perp}^{\vert l \vert } e^{il\Phi_{\text{c.m.}}}e^{ikZ_{\text{c.m.}}}\cos\theta\right\rbrace + \text{h.c.}.
\end{align}
The dot products of the type $ \textbf{r}'\cdot\textbf{E}_0 $ are replaced by $ r'\sqrt{4\pi /3}\sum_{\sigma =0,\pm 1}\epsilon_{\sigma}Y_1^{\sigma}(\theta',\phi') $ with $ \epsilon_{\pm 1}=(E_x\pm iE_y)/\sqrt{2} $ and $ \epsilon_0 = E_z $. In paraxial approximation, the $E_z  $ component is negligible. Equation (6) shows that within electric dipole approximation the polarization of the field interacts with the electronic motion and the field OAM interacts only with the external c.m. motion as also demonstrated by several authors \cite{enk_94, romero_02, babiker_02, alex_05}. The first term in Eq. (7) implies that in electric quadrupole transition, the field OAM is coupled to the c.m. motion only and the extra unit of angular momentum in electronic motion results from the quantized c.m. motion of the atom as in Ref. \cite{enk_94}. The invariance of the interaction Hamiltonian around the beam axis imposes the conservation of total angular momentum of the field plus atom system while the gradient of the field along radial direction couples the quantized c.m. and electronic motion. This coupling of c.m. and electronic motion is the main novel feature of this interaction. Our calculation shows that either of the terms $(\vert l\vert +l)  $ or $ (\vert l\vert -l) $ is nonzero depending on the sign of $ l $. The quadrupole transition matrix element is given by $ \mathcal{M}_{i\rightarrow f}^q=\langle \Upsilon_f\vert H_{\text{I,atom}}^{q}\vert \Upsilon_i\rangle $, where $ \Upsilon $ denotes an unperturbed atomic state, i.e., eigenstate of $ H_0 $. We assume $ \Upsilon (\textbf{R}_{\text{c.m.}},\textbf{r}) = \Psi_{\text{c.m.}}(\textbf{R}_{\text{c.m.}})\psi (\textbf{r})   $, where the c.m. wavefunction  $ \Psi_{\text{c.m.}}(\textbf{R}_{\text{c.m.}}) $ depends on the external potential that traps the atom and the internal electronic wavefunction $ \psi (\textbf{r}) $ can be considered to be a highly correlated Coupled-Cluster orbital \cite{mondal_13}.

\begin{align}
\mathcal{M}_{i\rightarrow f}^q&=\frac{1}{2}\sqrt{\dfrac{4\pi }{3\vert l\vert !}} e\left(\frac{m_n}{m_t}\right)^2\sum_{\sigma =0,\pm1}\epsilon_{\sigma}\left\lbrace \vert l\vert \left(\dfrac{w_{\text{c.m.}}}{w_0}\right)^{\vert l\vert-1}\left(\dfrac{w_e}{w_0}\right)\langle \psi_f\vert \frac{r^2}{w_e} Y_1^{\sigma}({\bf\hat{\textbf{r}}})\sin\theta e^{\text{sgn}(l)i\phi}\vert \psi_i\rangle \mathcal{M}_{\text{c.m.}}^{\text{sgn}(l)(\vert l\vert-1)} \right. \nonumber\\
&\left. + (ik)\left(\dfrac{w_{\text{c.m.}}}{w_0}\right)^{\vert l\vert}  \langle \psi_f\vert r^2Y_1^{\sigma}({\bf\hat{\textbf{r}}})\cos\theta \vert \psi_i\rangle \mathcal{M}_{\text{c.m.}}^{l}\right\rbrace
\end{align}
where $ w_{\text{c.m.}} $ and $ w_e $ stand for the average spatial width of $ \Psi_{\text{c.m.}}(\textbf{R}_{\text{c.m.}}) $ and $ \psi (\textbf{r}) $, respectively and $ \mathcal{M}_{\text{c.m.}}^{l'}= \langle \Psi_{\text{c.m.}f}\vert \left(\dfrac{\text{R}_{\text{c.m.}\perp}}{w_{\text{c.m.}}}\right)^{\vert l' \vert } e^{il'\Phi_{\text{c.m.}}}e^{ikZ_{\text{c.m.}}}\vert \Psi_{\text{c.m.}i}\rangle $.
\end{widetext} 
From Eq. (8) we deduce the selection rule for magnetic quantum number $ \Delta m=0, \pm 1, \pm 2 $. The extra unit of angular momentum transferred to the electron comes from the quantized c.m. motion of the atom. The orientation of this angular momentum is the same as that of the field OAM. We have assumed that $ w_{\text{c.m.}}\gg w_e $. Hence, the transition probabilities become insignificant for higher order multipole transitions unless the intensity of beam is very high.

\paragraph{\textbf{The molecule-radiation interaction:}}

We consider a diatomic molecular ion, e.g., H$_2^+$ or HD$ ^{+} $ for simplicity, comprising of three particles: Two nuclei of mass (charge) $ m_1 $ ($ +e $) and $ m_2 $ ($ +e $), and one electron $ m_e $ ($ -e $) as schematically shown in Fig.1. A and B are two nuclei and $ O' $ is the center of mass of the molecule. $ \textbf{r}_1, \textbf{r}_2 $ are the positions of the nuclei A and B with respect to the c.m., respectively. $ \textbf{r}_e $ is the position of the electron in c.m. coordinate system. A molecular axis is defined as position of A with respect to B, i.e., $ \overrightarrow{\text{BA}}=\textbf{R} $. The nuclei oscillations $ q $ are written explicitly as $ \textbf{R} = \bar{\textbf{R}}+\textbf{v}q $, where $ \bar{\textbf{R}} $ is the equilibrium position of A with respect to B, $ \textbf{v} $ is a constant vector with the same direction as that of $ \textbf{R} $ and of magnitude given by the properties of the molecule. $ \textbf{r}_A $ and $ \textbf{r}_B $ are positions of the electron relative to the nuclei A and B, respectively. Hence, $ \textbf{r}_e=\textbf{r}_1+\textbf{r}_A $, $ \textbf{r}_e=\textbf{r}_2+\textbf{r}_B  $ and $ \textbf{r}_e = \dfrac{1}{2} \lbrace r {\bf{\hat{\textbf{R}}}}+(\textbf{r}_A+\textbf{r}_B)\rbrace  $, where $ r = \vert \textbf{r}_1\vert - \vert \textbf{r}_2\vert = \dfrac{R(m_2-m_1)}{m_2+m_1} $ and $ {\bf{\hat{\textbf{R}}}} $ is unit vector along $ \textbf{R} $. The unperturbed quantum state of the diatomic molecule is described by $ \Psi  = \psi_e (\textbf{r}_A,\textbf{r}_B ) \psi_v (q) \psi_r ({\bf{\hat{\textbf{R}}}})\Psi_{\text{c.m.}}(\textbf{R}_{\text{c.m.}}) $, where $ \psi_e (\textbf{r}_A,\textbf{r}_B ) $ stands for the electronic state, $ \psi_v (q) $ stands for the vibrational wavefunction, $\psi_r ({\bf{\hat{\textbf{R}}}}) $ stands for the rotational wavefunction, and $ \Psi_{\text{c.m.}}(\textbf{R}_{\text{c.m.}}) $ stands for the c.m. wavefunction.
\begin{figure}
\includegraphics[trim= 3cm 2cm 0cm 2cm, scale=.5]{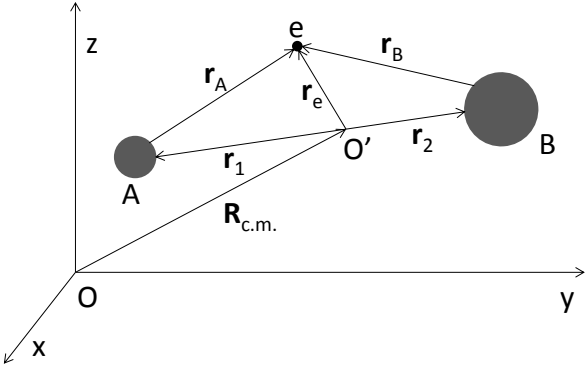}
\caption{Diatomic molecule in the laboratory coordinate system. $ z $ axis is along the direction of propagation of the LG beam.}
\end{figure}
The interaction Hamiltonian is derived in PZW scheme as in the previous section. Again, we assume that molecular dimension is so small compared to the beam waist ($ \vert \textbf{r}_e\vert \ll \vert \textbf{R}_{\text{c.m.}}\vert $) and use Taylor's expansion about c.m. position,

\begin{widetext}
\begin{align}
\textbf{E}\left( \textbf{R}_{\text{c.m.}}+\lambda \textbf{r}_e\right)= \textbf{E}( \textbf{R}_{\text{c.m.}})+\dfrac{\textbf{E}_0}{\sqrt{\vert l\vert !}}\lambda \left(\dfrac{1}{w_0}\right)^{\vert l\vert} r_e\left\lbrace {\bf\hat{\textbf{r}}}_e\cdot \vec{\nabla} \left({\text{r}'}_{\perp}^{\vert l \vert } e^{il\phi'}e^{ikz'}\right)\right\rbrace_{\textbf{R}{\text{c.m.}}}+\cdots ,
\end{align}
where
\begin{align}
r_e\left\lbrace {\bf\hat{\textbf{r}}}_e\cdot \vec{\nabla} \left({\text{r}'}_{\perp}^{\vert l \vert } e^{il\phi'}e^{ikz'}\right)\right\rbrace_{\textbf{R}{\text{c.m.}}}=-2l \text{R}_{\text{c.m.}\perp}^{\vert l \vert -1 } e^{\text{sgn}(l)i(\vert l\vert -1)\Phi_{\text{c.m.}}}e^{ ikZ_{\text{c.m.}}}\mathcal{R}_1^{\text{sgn}(l)1}(\textbf{r}_e)+(ik)\text{R}_{\text{c.m.}\perp}^{\vert l \vert} e^{il\Phi_{\text{c.m.}}}e^{ikZ_{\text{c.m.}}}\mathcal{R}_1^{0}(\textbf{r}_e)
\end{align}
and $ \mathcal{R}_l^m(\textbf{r}') $ is regular solid spherical harmonic function given as $ \mathcal{R}_l^m(\textbf{r}')=\mathcal{N}_{l,m}(r')^lY_l^m(\theta',\phi') $ \cite{alex_05, alex_06}. The normalization constant is $ \mathcal{N}_{l,m}=\sqrt{4\pi/(2l+1)(l+m)!(l-m)!} $. Using the expression of $ \textbf{r}_e $ and translation property of solid spherical harmonics \cite{alex_06, rury_13}, Eq. (10) can be expressed as
\begin{align}
&\textbf{r}_e\cdot \left\lbrace\vec{\nabla} \left({\text{r}'}_{\perp}^{\vert l \vert } e^{il\phi'}e^{ikz'}\right)\right\rbrace_{\textbf{R}{\text{c.m.}}} \nonumber\\
&=-l \text{R}_{\text{c.m.}\perp}^{\vert l \vert -1 } e^{\text{sgn}(l)i(\vert l\vert -1)\Phi_{\text{c.m.}}}e^{ ikZ_{\text{c.m.}}}\left[\mathcal{R}_1^{\text{sgn}(l)1}(r {\bf{\hat{\textbf{R}}}})+\mathcal{R}_1^{\text{sgn}(l)1}(\textbf{r}_A+\textbf{r}_B)\right]\nonumber\\
&+(ik)(1/2)\text{R}_{\text{c.m.}\perp}^{\vert l \vert} e^{il\Phi_{\text{c.m.}}}e^{ ikZ_{\text{c.m.}}}\left[\mathcal{R}_1^{0}(r {\bf{\hat{\textbf{R}}}})+\mathcal{R}_1^{0}(\textbf{r}_A+\textbf{r}_B)\right].
\end{align}
%\end{widetext}
The second term of Eq. (9) shows that one unit of angular momentum is transferred from external c.m. motion to the internal rotational and electronic motion of the molecule. If we retain only first two terms in Eq. (9) and carry out the integration over $ \lambda $, we get dipole interaction Hamiltonian as given by

\begin{align}
H_{\text{I,mol}}^d =&\frac{e}{2} \sqrt{\dfrac{4\pi }{3\vert l\vert !}}\left(\dfrac{1}{w_0}\right)^{\vert l\vert}\sum_{\sigma =0,\pm 1}\epsilon_{\sigma}\left\lbrace \text{R}_{\text{c.m.}\perp}^{\vert l \vert }e^{il\Phi_{\text{c.m.}}}e^{ikZ_{\text{c.m.}}} \left[(r_AY_1^{\sigma}({\bf\hat{\textbf{r}}}_A)+r_BY_1^{\sigma}({\bf\hat{\textbf{r}}}_B))(1+\dfrac{ik}{4}\mathcal{R}_1^0(r {\bf{\hat{\textbf{R}}}}))\right.\right.\nonumber\\
&\left.+\dfrac{ik}{4} rY_1^{\sigma}({\bf{\hat{\textbf{R}}}})\mathcal{R}_1^{0}(\textbf{r}_A+\textbf{r}_B) \right]-\frac{l}{2}\text{R}_{\text{c.m.}\perp}^{\vert l \vert -1}e^{\text{sgn}(l)i(\vert l\vert -1)\Phi_{\text{c.m.}}}e^{ikZ_{\text{c.m.}}}\left[rY_1^{\sigma}({\bf{\hat{\textbf{R}}}})\mathcal{R}_1^{\text{sgn}(l)1}(\textbf{r}_A+\textbf{r}_B)\right.\nonumber\\
&\left.\left.+(r_AY_1^{\sigma}({\bf\hat{\textbf{r}}}_A)+r_BY_1^{\sigma}({\bf\hat{\textbf{r}}}_B))\mathcal{R}_1^{\text{sgn}(l)1}(r {\bf{\hat{\textbf{R}}}})\right]\right\rbrace +\text{h.c.}.
\end{align}

The transition matrix element of the dipole interaction Hamiltonian (12) is 
\begin{align}
\mathcal{M}_{i\rightarrow f}&=\langle \Psi_f \vert H_{\text{I, mol}}^d\vert \Psi_i\rangle\nonumber\\
&= \frac{e}{2}\sqrt{\dfrac{4\pi }{3\vert l\vert !}}\sum_{\sigma =0,\pm 1}\epsilon_{\sigma }\left\lbrace \left(\dfrac{w_{\text{c.m.}}}{w_0}\right)^{\vert l\vert}\left[\langle \psi_{ef}\vert (r_AY_1^{\sigma}({\bf\hat{\textbf{r}}}_A)+r_BY_1^{\sigma}({\bf\hat{\textbf{r}}}_B))\vert \psi_{ei} \rangle \left(\mathcal{M}_{r0}^0\mathcal{M}_{v}^0+\frac{ik}{4}\sqrt{\frac{4\pi }{3}}\left(\dfrac{m_2-m_1}{m_2+m_1}\right)\right.\right.\right. \nonumber\\
&\left.\left.\times\bar{R}\mathcal{M}_{r1}^0\mathcal{M}_{v}^1\right)+\left(\dfrac{ik}{4}\sqrt{\frac{4\pi }{3}}\right)\left(\dfrac{m_2-m_1}{m_2+m_1}\right)\bar{R}\langle \psi_{ef}\vert (r_AY_1^0({\bf\hat{\textbf{r}}}_A)+r_BY_1^0({\bf\hat{\textbf{r}}}_B))\vert \psi_{ei}\rangle\mathcal{M}_{r1}^{\sigma}\mathcal{M}_{v}^1\right] \mathcal{M}_{\text{c.m.}}^{l }\nonumber\\
&-\frac{l}{2}\sqrt{\frac{2\pi }{3}}\left(\dfrac{m_2-m_1}{m_2+m_1}\right)\left(\dfrac{w_{\text{c.m.}}}{w_0}\right)^{\vert l\vert -1}\left(\dfrac{\bar{R}}{w_0}\right)\left[\langle \psi_{ef}\vert (r_AY_1^{\text{sgn}(l)1}({\bf\hat{\textbf{r}}}_A)+r_BY_1^{\text{sgn}(l)1}({\bf\hat{\textbf{r}}}_B))\vert \psi_{ei}\rangle  \mathcal{M}_{r1}^{\sigma} \mathcal{M}_{v}^1 \right.\nonumber\\
&\left.\left.+\langle \psi_{ef}\vert (r_AY_1^{\sigma}({\bf\hat{\textbf{r}}}_A)+r_BY_1^{\sigma}({\bf\hat{\textbf{r}}}_B))\vert \psi_{ei} \rangle \mathcal{M}_{r1}^{\text{sgn}(l)1} \mathcal{M}_{v}^1 \right]\mathcal{M}_{\text{c.m.}}^{\text{sgn}(l)(\vert l\vert -1)}\right\rbrace,
\end{align}
where $ \mathcal{M}_{\text{c.m.}}^{l'}= \langle \Psi_{\text{c.m.}f}\vert \left(\dfrac{\text{R}_{\text{c.m.}\perp}}{w_{\text{c.m.}}}\right)^{\vert l' \vert } e^{il'\Phi_{\text{c.m.}}}e^{ikZ_{\text{c.m.}}}\vert \Psi_{\text{c.m.}i}\rangle $, $ \mathcal{M}_{rl}^{m}=\langle \psi_{rf}\vert Y_l^m(({\bf{\hat{\textbf{R}}}}))\vert \psi_{ri}\rangle $ and $ \mathcal{M}_{v}^p = \sum_{s=0}^p \dbinom{p}{s}(v/\bar{R})^s\langle \psi_{vf}\vert q^s\vert \psi_{vi}\rangle $.
\end{widetext}
It is clear from Eq. (12) and (13) that the c.m. motion can couple with the internal motion in interaction of LG beam with diatomic molecule even within electric dipole approximation. In case of homonuclear (non-polar) molecules, $ r=0 $. Thus Eq. (12) clearly shows that rotational and vibrational motion of non-polar molecules can not be influenced by OAM of LG beam. However, when non-polar diatomic molecules are polarized and adiabatically or non-adiabatically aligned with a non-resonant intense laser field \cite{friedrich_95, seideman_95, Lemeshko_13}, then it is possible for the aligned molecules to interact with a LG beam, leading to transfer of light OAM to the rotational motion of the molecules as we elaborate below.

Particularly interesting prospect for light OAM transfer to molecular electronic motion opens if one makes use of the rotational confinement due to the alignment of internuclear axes of diatomic molecules with a linearly polarized  off-resonant intense laser field \cite{friedrich_95, seideman_95, Lemeshko_13}. The alignment and orientation of diatomic molecules and consequent phenomena of anisotropic ionization \cite{spanner_12} and high harmonic generation (HHG) \cite{courtial_97, kanai_05, ramakrishna_07},  have now become an important topic of research  both from experimental and theoretical  points of view, providing new insight into the underlying effects of rotational confinement on molecular electronic orbitals. Since, the rotation of internuclear axis of a diatomic molecule can couple to the molecular electronic orbital, the interaction of aligned molecules with light carrying OAM will be interesting for OAM transfer in dipole interactions. The alignment of nonpolar ground state molecules is basically a result of two-photon off-resonant dipole interaction of molecules. The major advantage of using pre-aligned molecules would be to get a preferential direction of orientation for angular momentum transfer. If a far-off-resonant and  relatively intense  LG beam is allowed to interact with pre-aligned molecules, it is possible to orient the axis of the molecules by the transfer  of light orbital angular momentum. Since the molecular axis orientation is coupled to the  electronic angular  momentum, light OAM will eventually alter the electronic angular momentum.  Thus, our theoretically proposed mechanism of light OAM transfer can be experimentally verified by applying an LG beam to  diatomic molecules which are already aligned and subsequently probing  rotational dynamics with a probe laser via detecting ionization and HHG signals \cite{kanai_05}. The light OAM transfer can be inferred by comparing  results with and without LG beam. Thus, with the currently available technology of molecular alignment and orientation with pump-probe type setup, it is  possible to manipulate molecular rotational and electronic orbital motion with light carrying orbital angular momentum. 

In the next section, we discuss numerical results showing the effect of the coupling of c.m. and electronic motions on the quadrupole transition rate in interaction of a LG beam with $ ^{23} $Na BEC.

\section{Quadrupole interaction of a BEC with an LG beam: numerical results}
In the previous section, we have shown that the quadrupole transition is the lowest order transition in interaction of cold atoms with an LG beam for transferring field OAM to electronic motion via quantized c.m. motion. Here we present numerical results showing the effect of the coupling between c.m. and internal motion on quadrupole transition rates in interaction of an atomic BEC with an LG beam. The first term of Eq. (8) shows this transfer of optical OAM to electronic motion via quantized c.m. motion. This term is important to reveal the predicted effect in quadrupole transitions. 

Here we calculate the quadrupole transition rates considering quantum mechanical and coupled motion of both electronic and c.m. degrees of freedom of atoms. We consider the c.m. atomic state as the ground state and different vortex states of a trapped BEC \cite{andersen_06, tempere_01, dalfovo_96, edwards_96}. The system is dilute enough so that the Gross-Pitaevskii (GP) theory \cite{stringari_03} for trapped bosons is applicable to the c.m. motion. Vortex states in BEC will be created due to transfer of OAM from the LG beam to the c.m. motion of the atoms. In addition to the OAM, linear momentum (LM) of light will also be transferred to the c.m. motion of atoms. Hence, we write the initial and final stationary state of c.m. motion of atoms as \cite{dalfovo_96}
\begin{align}
\Psi_{\text{c.m.}i}(\textbf{R}_{\text{c.m.}})=\psi_{\text{c.m.}i}(\text{R}_{\text{c.m.}\perp},Z_{\text{c.m.}})e^{i\kappa_i \Phi_{\text{c.m.}}}
\end{align}
and
\begin{align}
\Psi_{\text{c.m.}f}(\textbf{R}_{\text{c.m.}})=\psi_{\text{c.m.}f}(\text{R}_{\text{c.m.}\perp},Z_{\text{c.m.}})e^{i\kappa_f \Phi_{\text{c.m.}}}e^{ikZ_{\text{c.m.}}}
\end{align}
where $ \kappa $ is the quantum of circulation of atoms about $ z $ axis. $ \kappa \neq 0 $ represents vortex states of the BEC. The expression of $ \mathcal{M}_{\text{c.m.}}^{l'} $ gives the selection rule for c.m. motion as $ \Delta \kappa = l' $. According to time-independent GP theory, the nonlinear Schr\"{o}dinger equation obeyed by the c.m. motion in an anisotropic simple harmonic potential trap is given by \cite{dalfovo_96, stringari_03}
\begin{widetext}
\begin{align}
\left[-\dfrac{\hbar^2}{2m_t}\left(\dfrac{\partial^2}{\partial \text{R}_{\text{c.m.}\perp}^2} +\kappa^2\text{R}_{\text{c.m.}\perp}^{-2}+\dfrac{\partial^2}{\partial Z_{\text{c.m.}}^2}\right)+\dfrac{m_t}{2}\left(\omega_{\perp}^2\text{R}_{\text{c.m.}\perp}^{2}+\omega_Z^2Z_{\text{c.m.}}^2\right)+\dfrac{4\pi \hbar^2 a}{m_t}\vert \psi_{\text{c.m.}}\vert^2 \right]\psi_{\text{c.m.}}=\mu \psi_{\text{c.m.}}
\end{align}
\end{widetext}
where $ \omega_{\perp} $ and $ \omega_Z $ are the two angular frequencies associated with the external potential of the anisotropic trap, $ a $ is the $ s $-wave scattering length, and $ \mu $ is the chemical potential. In Eq. (8), $ w_{\text{c.m.}} $ is of the order of the characteristic length of trap $ a_{\perp}=\left(\dfrac{\hbar}{m_t\omega_{\perp}}\right)^{1/2} $ and $ w_e $ is of the order of Bohr radius $ a_0 $. We evaluate the c.m. wavefunction at zero temperature using the steepest descent method for functional minimization as prescribed in Ref. \cite{dalfovo_96}. The electronic portion of the transition matrix element is calculated using Coupled-Cluster theory \cite{mondal_13}. 

We now proceed to numerically evaluate the quadrupole transition rates considering the first term of Eq. (8) where the c.m. and electronic motions are coupled. Let us consider a left circularly polarized LG beam ($ \sigma = +1 $ in Eqs. (6-8)) with $ l=+2 $ is interacting with a BEC of $ 10^3 $ number of $ ^{23} $Na atoms in an anisotropic harmonic trap. The axis of the beam and the axis of the trap are same and along z axis of the laboratory frame. Here it is important to note that, one unit of the field OAM changes the vorticity of the c.m. wavefunction and the other unit of field OAM is transferred to electronic motion via quantized c.m. motion.

For numerical illustration, we choose the characteristics of the experimental trap as given in Ref. \cite{andersen_06}. The asymmetry parameter of the trap is $ \lambda_{tr} = \omega_Z/\omega_{\perp}=2 $. The axial frequency $ \omega_Z/2\pi = 40 $ Hz. The corresponding characteristic length is $ a_{\perp}= 4.673\times 10^{-6} $ m. $ s $- wave scattering length $ a=2.75 $ nm \cite{simula_08, inouye_98}. The waist of the LG beam is $ w_0=10^{-4} $ m and the intensity $ I=10^2 $ Wcm$ ^{-2} $. The amplitude of the LG beam in Eq. (8) is related to the intensity by $ I= \epsilon_0c\epsilon_1^2/2 $ where $ \epsilon_0 $ is vacuum permittivity.

The initial electronic state is $ \vert 3S_{\frac{1}{2},\frac{1}{2}}\rangle $. The electronic portion $ \langle \psi_f\vert r^2Y_1^{\sigma}({\bf\hat{\textbf{r}}})\cos\theta \vert \psi_i\rangle $ of the second term in the R.H.S. of Eq. (8) indicates that the magnetic quantum number of final electronic state $ \psi_f $ will be changed by one unit due to the polarization of light. So, the final electronic state will be $ \vert 3D_{\frac{3}{2},\frac{3}{2}}\rangle $ or $ \vert 3D_{\frac{5}{2},\frac{3}{2}}\rangle $. But the electronic portion $ \langle \psi_f\vert \frac{r^2}{w_e} Y_1^{\sigma}({\bf\hat{\textbf{r}}})\sin\theta e^{\text{sgn}(l)i\phi}\vert \psi_i\rangle $ of the first term on the R.H.S. of Eq. (8)  shows that, in addition to the polarization of the beam, one unit of field OAM can be transferred to the electronic motion via quantized c.m. motion. That increases the magnetic quantum number by two units in the final electronic state, i.e., $ \vert 3D_{\frac{5}{2},\frac{5}{2}}\rangle $. Thus, the electronic transition $\vert 3S_{\frac{1}{2},\frac{1}{2}}\rangle \longrightarrow \vert 3D_{\frac{5}{2},\frac{5}{2}}\rangle  $ can be realized. Next, we show  how quadrupole transition rates are enhanced due to quantized c.m. motion and its coupling with the internal motion.

Fig. 2. presents the radial part of the c.m. wavefunction of the cold atom corresponding to the ground state and different vortex states in the simple harmonic potential trap.
\begin{figure}
\includegraphics[trim = 1cm 0cm 0cm 0cm, scale=.3]{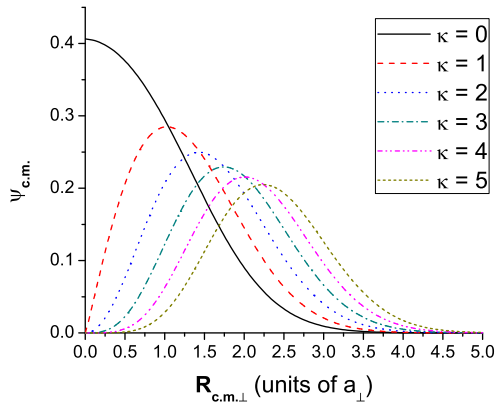}
\caption{(Color online) Radial dependence of the c.m. wavefunction corresponding to ground state and different vortex states of BEC of $ ^{23} $Na atoms.}
\end{figure}
The value of the electronic portion of the transition matrix element remains unchanged during variation of vorticity of the c.m. wavefunction. In ground state of c.m. motion, the probability of finding the atom on the beam axis is maximum. But as the vorticity of the states increases the probability of finding the atom away from beam axis increases. Hence, the value of the c.m. matrix element $ \mathcal{M}_{\text{c.m.}}^{1} $ increases. This in turn influences the electronic motion due to coupling with c.m. motion. Table I presents variation of the c.m. transition matrix element and the quadrupole transition rates ($  W^q $) with vorticity of involved c.m. states. It is clearly shown that with increase in $ \kappa_i $ the quadrupole transition rate increases due to increase in spread of the c.m. wavefunction. It is worth pointing out that $ W^q $ is enhanced because the c.m. motion is treated quantum mechanically and it is coupled with the electronic motion.
\begin{table}
\caption{Variation of $ \mathcal{M}_{\text{c.m.}}^{1} $ (in unit of $ a_{\perp} $) and quadrupole transition rates (in $ 10^3 $s$ ^{-1} $) with vorticity of initial state of c.m. wavefunction.}
\centering
\begin{tabular}{ccccccc}
\hline \hline
$ \kappa_i $&&$ \kappa_f $&&$ \mathcal{M}_{\text{c.m.}}^{1} $ && $ W^q $ \\
\hline
 0 && 1 &&1.00093&& 2.48\\
 1 && 2 &&1.41422&& 4.94\\
 2 && 3 &&1.73205&& 7.42\\
 3 && 4 &&2.00000&& 9.89\\
 4 && 5 &&2.23606&& 12.36\\
 5 && 6 &&2.44948&& 14.83\\
 \hline
\end{tabular}
\end{table}

Before ending this section, it is worthwhile to mention that, though for simplicity we have neglected electron's spin,  the theoretical treatment we have presented can be extended to include electron's spin, in particular in situations where spin-orbit interaction of atoms or molecules is strong. In that case and in absence of hyperfine interaction, the good quantum number of an atom or molecule will be ${\mathbf J}_e = \mathbf{L}_e + \mathbf{S}$, where $\mathbf{L}_e$ and $\mathbf{S}$ stand for total orbital and spin angular momentum of all the valence electrons of an atom or a molecule. In case of diatomic molecules, the projection $\Omega$ of $\mathbf{J}_e$ on the internuclear axis of the molecule will be a good quantum number. As in atoms and molecules, spin-orbit interaction also arises in light. Circular polarization of light is associated with the spin of a photon.  At a fundamental level, it is the vectorial sum of polarization  and orbital angular momenta of photon  which is a conserved quantity.  Light OAM will be nearly  a good quantum number under paraxial approximation only. In recent times several experiments \cite{friese_98, Piccirillo_02, nagali_09, Piccirillo_10, roy_13}  have demonstrated the transfer of angular momentum in light-matter interactions through spin-orbit interactions or inter-conversion between spin and angular momenta  of photons in different physical systems, for example,  in nematic liquid crystal \cite{Piccirillo_02, nagali_09, Piccirillo_10} and nanoplasmonics \cite{friese_98, roy_13}. The transfer of angular momentum in interaction of light with nematic liquid crystal is quite interesting. The mechanism of light orbital angular momentum transfer presented in this paper is consistent with the method of OAM transfer in liquid crystal as described in Ref.\cite{Piccirillo_02}.

\section{Conclusion}
In conclusion, we have studied the interaction of LG beam with an atom and a diatomic molecule. We have found that if the atom or molecule is cold enough for its c.m. motion to be quantized, then an angular momentum exchange can take place between c.m. and internal motion of the system. Atomic and molecular dimensions are far too small compared to the core size of the LG beam. However, if the atom or molecule is cooled down to its recoil limit such that the spread of its c.m. wavefunction is comparable to the wavelength of the beam, then it can feel the spatial variation of the electric field along the radial direction. The interaction with LG beam of Eq. (1) ensures that the orientation of the angular momentum transferred from c.m. to internal motion is same as that of the field OAM. Our calculations clearly show that the extra angular momentum (other than that coming from the polarization of the field) to internal motion is coming from the quantized c.m. motion. It may be possible to observe this effect with ultra-cold atoms or molecules interacting with LG beam as suggested by Van Enk \cite{enk_94} twenty years ago. We have numerically calculated the quadrupole transition rates in interaction of LG beam with atomic BEC where both the c.m. and electronic motions are quantized and coupled. We have shown the dependence of the transition rates on the vorticity of the c.m. wavefunction involved in the transition. In case of molecules at ultracold temperature or in quantum degenerate molecular gases, it would be possible to transfer the field OAM to internal motion even in dipole interaction. One potential application of the discussed effect can be thought of in quantum information processing using entangled angular momentum observables belonging to the same atom or molecule. Muthukrishnan and Stroud \cite{muthu_02} have shown the entanglement between electronic and c.m. degrees of freedom of a cold atom interacting with LG beam. They have considered electric dipole interaction where the field OAM couples to c.m. only and the field polarization couples to electronic motion. If the c.m. motion is allowed to couple with the internal degrees of freedom through laser-generated synthetic gauge field \cite{lin_11} then it is possible to realize transfer of light orbital angular momentum into the internal motion in otherwise similar experimental scenario of Ref. \cite{muthu_02}. This will provide a new avenue for entanglement manipulation in angular degrees-of-freedom. In this context, it is encouraging to note that the entanglement between electronic and vibrational degrees-of-freedom is already produced for an ion in a linear trap \cite{monroe_95} which can be a suitable system to explore the effect predicted in this paper. Our calculations further suggest that the quantized c.m. motion of an atom can couple to electronic motion in electric quadrupole interaction. The situation is even more interesting in case of molecule. Eq. (12) suggests that the c.m., electronic, rotational and vibrational motion can be coupled in interaction with LG beam even in electric dipole approximation. We have discussed how the induced polarization and alignment of diatomic molecules due to linearly polarized intense laser field can be useful for experimental verification of our theory. The alignment and orientation of the molecules of nematic liquid crystal have been found to play an important role in OAM transfer \cite{Piccirillo_02, nagali_09, Piccirillo_10, roy_13}, indicating that similar effects in gas phase molecules in intense laser fields will be important for light  OAM transfer to the molecules. Finally, the recent advent of ultra long-range cold Rydberg molecules whose electronic orbital can be as large as 100 nm \cite{greene_00, Bendkowsky_09} and the demonstration of the alignment of such molecules \cite{krupp_14} open a new avenue for studying OAM transfer in light-molecule interactions with huge enhancement of the size effects for the ability of the molecular electrons to experience  the spatial variation of light intensity.

Pradip Kumar Mondal acknowledges financial support from the Council of Scientific and Industrial Research (CSIR), India.

\end{document}